\documentclass[prb,aps,showpacs,superscriptaddress,twocolumn]{revtex4-1}
\usepackage{latexsym}
\usepackage{amsfonts}
\usepackage{amssymb}
\usepackage{amsmath}
\usepackage{graphicx}
\begin{document}
\title{Nano-scale strain engineering of graphene and graphene-based devices}
\date{\today}
\author{N.-C. Yeh,$^{1,2,3}$ C.-C. Hsu,$^{1,2}$ M. L. Teague,$^{1,2}$ J.-Q. Wang,$^4$ D. A. Boyd,$^1$ and C.-C. Chen$^{2,}$}

\affiliation{Department of Physics, California Institute of Technology, Pasadena, CA 91125, USA\\ 
$^2$Institute for Quantum Information and Matter, California Institute of Technology, Pasadena, CA 91125, USA\\ 
$^3$Kavli Nanoscience Institute, California Institute of Technology, Pasadena, CA 91125, USA\\
$^4$Department of Physics, Fudan University, Shanghai, China}

\begin{abstract}
Structural distortions in nano-materials can induce dramatic changes in their electronic properties. This situation is well manifested in graphene, a two-dimensional honeycomb structure of carbon atoms with only one atomic layer thickness. In particular, strained graphene can result in both charging effects and pseudo-magnetic fields, so that controlled strain on a perfect graphene lattice can be tailored to yield desirable electronic properties. Here we describe the theoretical foundation for strain-engineering of the electronic properties of graphene, and then provide experimental evidences for strain-induced pseudo-magnetic fields and charging effects in monolayer graphene. We further demonstrate the feasibility of nanoscale strain engineering for graphene-based devices by means of theoretical simulations and nano-fabrication technology.
\end{abstract}
\maketitle

\section{Introduction}

The advances of nano-science and technology have enabled new approaches to designing and controlling the properties of materials at the atomic and molecular length scales. Graphene, a single layer of carbon atoms forming a honeycomb lattice structure, exhibits many interesting properties~\cite{1,2,3} and have shown significant dependence of the electronic properties on the nanoscale structural distortion, largely because of the nano-material nature and the characteristics of the Dirac fermions,~\cite{4,5,6,7,8,9,10,11,12} which are massless, relativistic carriers of graphene with a linear energy-momentum dispersion relation.~\cite{1}  

In general, structural distortions in graphene may be associated with surface ripples, topological defects, adatoms, vacancies, and extended defects such as edges, cracks and grain boundaries.~\cite{1,4,5,6,7,8} The distortion-induced strain~\cite{9} in the graphene lattice typically gives rise to two primary effects on the Dirac fermions. One is an effective scattering scalar potential~\cite{4,5,6} and the other is an effective gauge potential.~\cite{7,8} For instance, compression and dilation distortion can lead to charging effects in localized regions,~\cite{1,4,5,6} which has been manifested as strain-enhanced local density of states by scanning tunneling microscopic and spectroscopic (STM/STS) studies of graphene~\cite{10} that was grown by means of the high-temperature chemical vapor deposition (CVD) technique.~\cite{13,14} It has also been theoretically proposed that certain controlled strain can induce a nearly uniform pseudo-magnetic field by lifting the valley degeneracy of graphene.~\cite{12,13} This prediction was first verified empirically via STM/STS studies of graphene nano-bubbles grown on Pt(111) substrates,~\cite{11} and subsequently reported in thermal CVD-grown graphene on copper.~\cite{10} The presence of significant pseudo-magnetic fields could lead to the localization of Dirac fermions, whereas the charging effect could lead to strong scattering of Dirac fermions, both contributing to the reduction of electrical mobility. Further, extended structural distortion may lead to long-range symmetry breaking, giving rise to fundamental changes in the electronic bandstructures such as gap opening in the Dirac spectra and spontaneous local time-reversal breaking.~\cite{12,15,16} The symmetry-breaking distortion may give rise to novel gauge potentials, such as certain non-Abelian gauge potentials~\cite{15} and the Kekule distortion,~\cite{16} yielding novel electronic states such as fractionally quantized energy spectra as seen in STS studies of strained graphene.~\cite{10,12}     

The susceptibility of graphene to structural distortions can in fact provide opportunities for engineering unique electronic properties of graphene.~\cite{17} For instance, the presence of pseudo-magnetic fields could be applied to the development of ``valleytronics''~\cite{18,19,20} by lifting the valley degeneracy of graphene. Proper nanoscale strain engineering may also produce tunable electronic density of states for nano-electronic devices such as electron beam collimators/deflectors or field effect transistors.~\cite{17,20} On the other hand, the feasibility of nanoscale strain engineering relies on the premise of subjecting an ideal graphene sheet to controlled structural distortion. While exfoliated graphene could achieve nearly ideal graphene characteristics, the extremely small sheets and the non-scalable approaches to the production are not compatible with strain-engineering of practical devices. Alternatively, chemical vapor deposition (CVD) techniques at high temperatures ($\sim 1000 ^{\circ}$C) have been shown to produce large sheets of graphene,~\cite{13} although the resulting graphene often reveals significant spatial strain variations~\cite{10} unless numerous additional processing steps were taken.~\cite{14} The presence of uncontrollable strain distribution would not be suitable for nanoscale strain-engineering. In this context, the capability of reproducibly growing sizable high-quality and strain-free graphene is necessary for realizing the concept of strain engineering. 

We have recently developed a new growth method based on plasma-enhanced chemical vapor deposition (PECVD) at reduced temperatures.~\cite{21} This new method has been shown to reproducibly achieve, in one step, high-mobility large-sheet graphene samples that are nearly strain free,~\cite{21} paving the way for realization of nanoscale strain-engineering. The objective of this work is to demonstrate the feasibility of nanoscale strain-engineering of graphene by means of both theoretical simulations and empirical development of PECVD-grown graphene on nanostructured substrates. The correlation between realistic designs of nanoscale strain distributions and the resulting effects on the electronic properties of graphene is evaluated by means of molecular dynamics. Empirical proof of concept for nanoscale strain engineering via nanostructured substrates for graphene is also demonstrated and compared with theoretical simulations. Finally, we discuss feasible device applications based on nanoscale strain-engineering of graphene.

\section{Effects of structural distortions on the electronic properties of graphene}

The physical causes for the strain-induced effective scalar and gauge potentials in graphene may be understood in terms of the changes in the distance or angles between the $p_z$ orbitals that modify the hopping energies between different lattice sites, thereby giving rise to the appearance of a vector (gauge) potential ${\cal A}$ and a scalar potential $\phi$ in the Dirac Hamiltonian.~\cite{1,7} Under the preservation of global time-reversal symmetry, the presence of strain-induced vector potential leads to opposite pseudo-magnetic fields $\textbf{B}_s = \nabla \times {\cal A}$ and $\textbf{B}_s ^{\ast} = \nabla \times {\cal A} ^{\ast} = - \nabla \times {\cal A} = - \textbf{B}_s$ for the two inequivalent Dirac cones at $K$ and $K^{\prime}$, respectively. On the other hand, the presence of a spatially varying scalar potential can result in local charging effects known as self-doping.~\cite{1} 

\subsection{Overview of the effect of strain on Dirac fermions of graphene}

As mentioned in the introduction, disorder generally induces two types of contributions to the original Dirac Hamiltonian ${\cal H} _0$. One is associated with the charging effect of a scalar potential, and the other is associated with a pseudo-magnetic field from a strain-induced gauge potential. For an ideal graphene sample near its charge-neutral point, the massless Dirac Hamiltonian ${\cal H} _0$ in the tight-binding approximation is given by:~\cite{1}
\begin{equation}
{\cal H}_0 = -t \sum _{\langle i,j \rangle,\sigma} (a ^{\dagger}_{i,\sigma} b_{j,\sigma} + H.c. ) - t^{\prime} \sum _{\langle i,j \rangle,\sigma} (a ^{\dagger}_{i,\sigma} a_{j,\sigma} + b ^{\dagger}_{i,\sigma} b_{j,\sigma} +H.c. ) ,
\label{eq:H0}
\end{equation}
where $a_{i,\sigma}$ ($a^{\dagger}_{i,\sigma}$) annihilates (creates) an electron with spin $\sigma$ ($\sigma = \uparrow , \downarrow$) on site $\textbf{R}_i$ of the sublattice A, $b_{i,\sigma}$ ($b^{\dagger}_{i,\sigma}$) annihilates (creates) an electron with spin $\sigma$ on site $\textbf{R}_i$ of the sublattice B, $t \approx 2.8$ eV ($t^{\prime} \approx 0.1$ eV) is the nearest-neighbor (next-nearest-neighbor) hopping energy for fermion hopping between different sublattices, and $H.c.$ refers to the Hermitian conjugate.~\cite{1} 

The presence of lattice distortion results in a spatially varying in-plane displacement field $\textbf{u} (x,y) = (u_x (x,y), u_y (x,y))$ and a spatially varying height displacement field $h(x,y)$. Therefore, in addition to changes in the distances or angles among the two-dimensional $\sigma$-bonds, distortion also induces changes in the distance or angle between the $p_z$ orbitals. Such three-dimensional lattice distortion gives rise to the following expressions for the strain tensor components $u_{ij}$ (where $i,j$ = $x,y$):~\cite{1} 
\begin{eqnarray}
u_{xx} &\equiv \frac{\partial u _x}{\partial x} + \frac{1}{2} \left( \frac{\partial h}{\partial x} \right) ^2, \quad \qquad \qquad \qquad \qquad \qquad \nonumber\\
u_{yy} &\equiv \frac{\partial u _y}{\partial y} + \frac{1}{2} \left( \frac{\partial h}{\partial y} \right) ^2, \quad \qquad \qquad \qquad \qquad \qquad \nonumber\\
u_{xy} &\equiv \frac{1}{2} \left( \frac{\partial u _x}{\partial y} + \frac{\partial u _y}{\partial x} \right) + \frac{1}{2} \left( \frac{\partial h}{\partial x} \frac{\partial h}{\partial y} \right), \qquad \qquad \qquad \quad
\label{eq:Strain}
\end{eqnarray}
so that an additional perturbative Hamiltonian ${\cal H} ^{\prime}$ must be introduced to describe the wave-functions of Dirac fermions in distorted graphene:~\cite{1,4,5}
\begin{eqnarray}
{\cal H} ^{\prime} &= \sum _{i,j,\sigma} \lbrack \delta t^{ab} (a ^{\dagger}_{i,\sigma} b_{j,\sigma} + H.c. ) + \delta t^{aa} (a ^{\dagger}_{i,\sigma} a_{j,\sigma} + b ^{\dagger}_{i,\sigma} b_{j,\sigma}) \rbrack \nonumber \\
 &= \int d^2 r \sum _{\alpha = 1,2} \lbrack \Phi (\textbf{r}) \Psi ^{\dagger} _{\alpha} (\textbf{r}) \Psi _{\alpha} (\textbf{r}) + \Psi ^{\dagger} _{\alpha} (\textbf{r}) (\tilde{\sigma} \cdot {\cal A} _{\alpha}) \Psi _{\alpha} (\textbf{r}) \rbrack \nonumber \\
 &\equiv {\cal H} _{\Phi} + {\cal H} _{\cal A} , \qquad \qquad \qquad \qquad \qquad \qquad \qquad \qquad
\label{eq:Hprime}
\end{eqnarray}
where $\delta t^{ab}$ and $\delta t^{aa} (= \delta t^{bb})$ are strain-induced modifications to the nearest and next nearest neighbor hopping energies, respectively, $\tilde{\sigma}$ represents the ($2 \times 2$) Pauli matrices, $\Psi _{\alpha} (\textbf{r})$ denotes the spinor operator for the two sublattices in the continuum limit, and ${\cal H} _{\Phi}$ (${\cal H} _{\cal A}$) represents the strain-induced perturbation to the scalar (gauge) potential. We note that in the event of finite $z$-axis corrugation, the modified hopping integrals for Dirac fermions from one site to another must involve not only the in-plane orbitals but also the $p_z$ orbitals of the distorted graphene sheet as the result of the three-dimensional distribution of electronic wavefunctions.  

Specifically, the strain-induced gauge potential ${\cal A} = (A_x, A_y)$ in Eq.~(\ref{eq:Hprime}) is related to the two-dimensional strain field $u_{ij}(x,y)$ by the following relation (with the x-axis chosen along the zigzag direction):~\cite{4,5} 
\begin{eqnarray}
A_x &= \pm \frac{\beta}{a_0} (u_{xx} - u_{yy}), \qquad \qquad \qquad \nonumber\\
A_y &= \mp 2 \frac{\beta}{a_0} u_{xy}. \qquad \qquad \qquad \qquad \quad
\label{eq:Gauge}
\end{eqnarray}
where $a_0 \approx 0.142$ nm is the nearest carbon-carbon distance, and $\beta$ is a constant ranging from 2 to 3 in units of the flux quantum~\cite{1}. Similarly, the compression/dilation components of the strain can result in an effective scalar potential $\Phi (x,y)$ in addition to the aforementioned pseudo-magnetic field so that there may be a static charging effect~\cite{4,5,6}. Here $\Phi (x,y)$ is given by:~\cite{4,5,6} 
\begin{equation}
\Phi (x,y) = \Phi_0 (u_{xx} + u_{yy}) \equiv \Phi_0 \bar{u},
\label{eq:phi}
\end{equation}
where $\Phi _0 \approx 3$ eV, and $\bar{u}$ is the dilation/compression strain. 

By inserting the explicit expressions given in Eqs.~(\ref{eq:Strain}), (\ref{eq:Gauge}) and (\ref{eq:phi}) into Eq.~(\ref{eq:Hprime}), we find that the excess strain components associated with the changes in the distances and angles of the $p_z$-orbitals may be written in forms of $\lbrack (\partial h/\partial x)^2 + (\partial h/\partial y)^2 \rbrack$ and $(\partial h/\partial x)(\partial h/\partial y)$, respectively. We further note that in the event of strong $z$-axis corrugation, the strain components resulting from the variations in the $p_z$-orbitals may become dominant over the in-plane strain components according to Eq.~(\ref{eq:Strain}). 

Thus, the total Hamiltonian for the Dirac fermions of graphene becomes ${\cal H} = {\cal H} _0 + {\cal H} ^{\prime} = {\cal H} _0 + ({\cal H} _{\Phi} + {\cal H} _{\cal A})$. The physical significance of the perturbed term ${\cal H} _{\cal A}$ is analogous to the contribution of a vector potential ${\cal A}$ in a two-dimensional electron gas that gives rise to a vertical magnetic field $\textbf{B}_s = \nabla \times {\cal A} = B_s \hat{z}$ and the formation of quantized orbitals. In this context, the Landau levels $E_n$ of Dirac fermions under a given $B_s$ satisfy the following relation:~\cite{1} 
\begin{equation}
E_n = {\rm sgn}(n) \sqrt{(2e v_{F} ^2 \hbar B_s)|n|},
\label{eq:En}
\end{equation}
where $v_F$ denotes the Fermi velocity of graphene and $n$ denotes integers. Additionally, the magnetic length $\ell _B$ associated with a given $B_s$ is $\ell _B \equiv \sqrt{(\hbar/eB_s)} \sim 25 \ \rm nm/\sqrt{B_s}$ ($B_s$ in units of Tesla).~\cite{1} Similarly, the physical significance of the perturbed term ${\cal H} _{\Phi}$ is analogous to the contribution from a scalar potential $\Phi$. For strain primarily induced by corrugations along the out-of-plane ($z$) direction, the typical strain is of order $\bar{u} \sim (\Delta z/L)^2$ from Eq.~(\ref{eq:Strain}), where $\Delta z$ denotes the height fluctuations and $L$ is the size of the lateral strained region. In principle the charging effect associated with the scalar potential can be suppressed in a graphene layer suspended over a metal if $\Delta z \ll \ell _B$.~\cite{7,8} However, for $\Delta z \sim \ell _B$, the strain-induced charging effect can no longer be effectively screened so that experimental observation of such an effect at the nanoscale may be expected.~\cite{10,12}  

\subsection{Empirical manifestation of giant pseudo-magnetic fields and local charging effects in strained graphene by STM/STS}

Significant charging effects and giant pseudo-magnetic fields in nanoscale strained graphene have been manifested by STM/STS studies of thermal CVD-grown graphene.~\cite{10,11,12} As exemplified in Fig.~1a for atomically resolved topography of a CVD-grown graphene on copper, significant lattice distortion along a ridge line is clearly visible. Using Eq.~(\ref{eq:phi}), the corresponding compression/dilation strain map obtained from the lattice distortions in Fig.~1a is shown in Fig.~1b, and the empirically determined zero-bias conductance map of the same area is shown in Fig.~1c. We find strong correlation among the maps of topography, strain and the tunneling conductance.     

\begin{figure}
\centering
\includegraphics[width=3.4in]{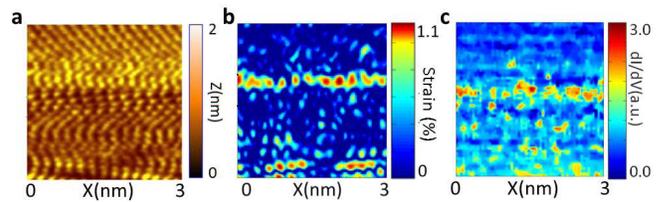}
\caption{Strain-induced charging effects on thermal CVD-grown graphene~\cite{10}: {\bf (a)} Atomically resolved STM topography over a $(3 times 3)$ nm$^2$ area, showing significantly distorted atomic structures. {\bf (b)}  The magnitude of biaxial strain obtained from using Eq.~(\ref{eq:phi}) and the displacement fields obtained from the topography in (a). {\bf (c)} The conductance map at $V = 0$ over the same area shown in (a). There is apparently strong correlation between the strain map in (b) and the conductance map in (c), which is consistent with strain-induced charging effects.}
\label{fig1}
\end{figure}

Empirically, the pseudo-magnetic fields $B_s$ can be determined from the strain-induced quantized conductance peaks superposed onto of the Dirac spectrum~\cite{7,8,10,11} according to Eq.~(\ref{eq:En}). As exemplified in Fig.~2, the tunneling conductance of strained graphene revealed quantized spectral peaks in addition to the generic Dirac tunneling spectrum. By plotting the peak positions in energy relative to the Dirac point $(E-E_{Dirac})$ versus $\sqrt{|n|}$, where $n$ being integers or fractional numbers ($p/q$) with $q = 3$ and $p$ being an integer,~\cite{10,12} we could identify the magnitude of $B_s$ by using Eq.~(\ref{eq:En}). From Fig.~2, we note that the pseudo-magnetic fields exhibited spatial variations depending on the local strain, and larger strain generally induced stronger $B_s$ values. The strain-induced giant pseudo-magnetic fields, on the order of tens of Tesla, have the effect of confining the motion of Dirac fermions to the magnetic length $\ell _B$, similar to the formation of electronic cyclotron orbit under an external magnetic field. Hence, the mobility of Dirac fermions becomes substantially compromised under the presence of giant pseudo-magnetic fields.   

\begin{figure}
\centering
\includegraphics[width=3.4in]{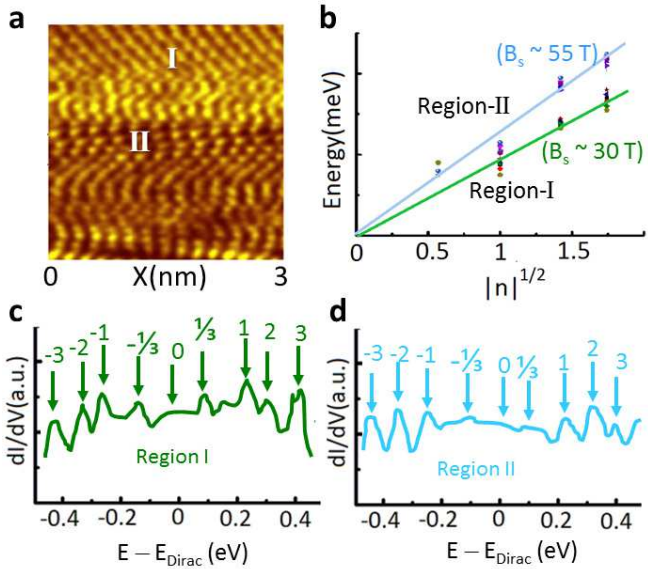}
\caption{Strain-induced giant pseudo-magnetic fields in thermal CVD-grown graphene:~\cite{10} {\bf (a)} Atomically resolved STM topography over a $(3 times 3)$ nm$^2$ area same as that in Fig.~1(a), with two distinctly different structural distortions denoted as Region-I and Region-II. {\bf (b)} A representative tunneling conductance of Region-I after subtraction of the standard Dirac spectrum, showing conductance peaks at quantized energies that correspond to the presence of a local   pseudo-magnetic field. {\bf (c)} A representative tunneling conductance of Region-II after subtraction of the standard Dirac spectrum, showing conductance peaks at quantized energies that correspond to the presence of a local pseudo-magnetic field larger than that in Region-I. {\bf (d)} Relation between $(E_n - E_{Dirac})$, the quantized energies relative to the Dirac energy $E_{Dirac}$, and $|n|^{1/2}$ in Regions-I and II. The slope corresponds to pseudo-magnetic fields $\sim 30$ and $\sim 55$ Tesla for Regions-I and II, respectively.}
\label{fig2}
\end{figure}

\subsection{Empirical manifestation of macroscopic strain effects by Raman spectroscopy}

The structural distortion-induced strain effects on graphene can also be determined by examining the Raman spectroscopy.~\cite{22,23} Specifically, the biaxial strain $(\epsilon _{ll} + \epsilon _{tt}) \approx (u_{xx} + u_{yy})$ in the PECVD-graphene on Cu can be estimated by considering the Raman frequency shifts $\Delta \omega _m \equiv (\omega _m - \omega _m ^0)$ and the Gruneisen parameter $\gamma _m ^{\rm biax}$:~\cite{22,23}  
\begin{equation}
\gamma _m ^{\rm biax} = \frac{\Delta \omega _m}{\omega _m ^0 (\epsilon _{ll} + \epsilon _{tt})},
\label{eq:gamma}
\end{equation}
where $m$ (= $G$, 2$D$) refers to the specific Raman mode.~\cite{22,23} Using $\gamma _{2D} ^{\rm biax} = 2.7$ and $\gamma _G ^{\rm biax} = 1.8$, we can determine the average strain of graphene samples prepared under different conditions. As exemplified in Fig.~3a-d for the comparison of the Raman spectroscopy taken on a thermal CVD-grown graphene and a low-temperature PECVD-grown graphene, we found a general trend of downshifted G-band and 2D-band frequencies for all PECVD-grown graphene relative to thermal CVD-grown graphene on the same substrate,~\cite{21} indicating reduced strain in PECVD-grown graphene. 

\begin{figure}
\centering
\includegraphics[width=3.4in]{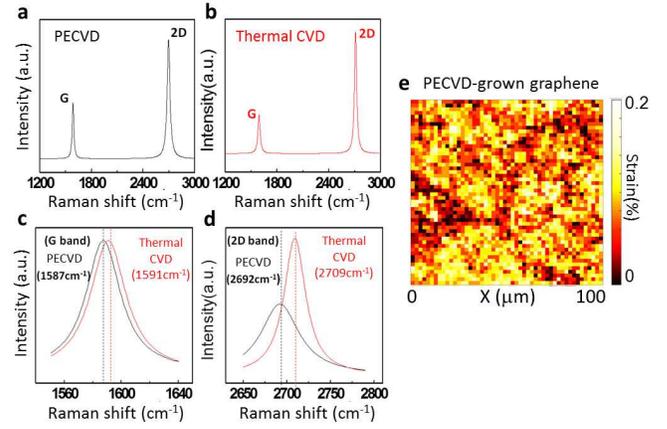}
\caption{Determining the strain effects in graphene from Raman spectroscopy:~\cite{21} {\bf (a)} A representative Raman spectrum (laser wavelength 514 nm) of a PECVD-grown graphene sample, showing sharp G- and 2D-bands that are characteristic of pure monolayer graphene. {\bf (b)} A representative Raman spectrum of a thermal CVD-grown graphene sample, also showing sharp G- and 2D-bands. {\bf (c)} Comparison of the G-band frequency of the PECVD- and thermal CVD-grown graphene, showing a downshifted G-band and therefore reduced strain in the PECVD-grown graphene. {\bf (d)} Comparison of the 2D-band frequency of the PECVD- and thermal CVD-grown graphene, showing a downshifted 2D-band and therefore reduced strain in the PECVD-grown graphene, consistent with the finding shown in (c) for the G-band. {\bf (e)} A $(100 \times 100) \mu m^2$ areal map of the magnitude of the biaxial strain of a PECVD-grown graphene sample, obtained from Raman spectroscopic measurements with Renishaw InVia at a laser wavelength 532 nm and a spatial resolution of 2 $\mu$m per pixel. The average biaxial strain for this area is $\sim 0.07$\%.}  
\label{fig3}
\end{figure}

In general, using Eq.~(\ref{eq:gamma}) we can determine the spatial strain distribution in a graphene sample over a substantially large area, as illustrated in Fig.~3e for a low-temperature PECVD-grown graphene sample over an area of $(100 \times 100) \ \mu m^2$ that appears to be nearly strain-free with an average of 0.07\% strain. This is in sharp contrast to the significant strain-induced effects found in thermal CVD-grown graphene. As further illustrated in Fig.~4 for both atomically resolved topographic studies by means of STM and for macroscopic-scale studies over multiple areas per sample and for multiple samples by means of Raman spectroscopy, we found that for the same Cu-foil substrates, the average strain in PECVD-grown graphene was consistently more than one order of magnitude smaller than that of our typical thermal CVD-grown graphene. 

While the magnitude of strain derived from Raman spectroscopy with Eq.~(\ref{eq:gamma}) does not contain the shear strain components, we remark that the information thus derived is still largely proportional to the total strain obtained from the displacement fields ($e.g.$, measured by STM) accordingly to Eq.~(\ref{eq:Strain}) if the symmetry between $x$ and $y$ components is preserved and the $z$-axis corrugation dominates. This is the reason why the empirical strain distributions shown in Fig.~4, which were obtained from both STM and Raman spectroscopic studies, are mostly consistent.  

Our finding of much reduced strain is also consistent with the much better electrical mobility in the PECVD-grown graphene, typically $40,000 \sim 70,000 \ \rm cm^2/V-s$ at 300 K, which is more than one order of magnitude better than that of thermal CVD-grown graphene.~\cite{21} The availability of nearly strain-free graphene by means of our PECVD growth method clearly paves the way to strain-engineering of the electronic properties of graphene. 

\begin{figure}
\centering
\includegraphics[width=3.4in]{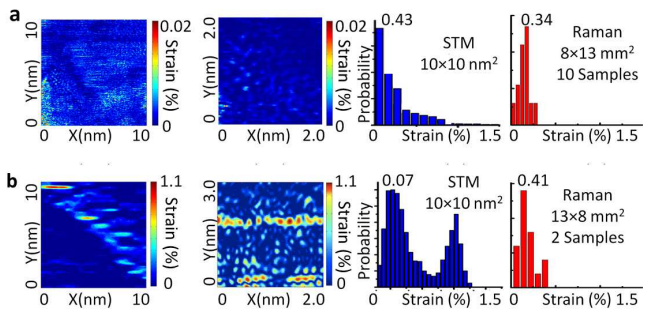}
\caption{Comparison of the spatially resolved strain maps and strain histograms of graphene grown by PECVD and thermal CVD methods from STM and Raman spectroscopic studies:~\cite{21} {\bf (a)} From left to right, compression/dilation strain maps over successively decreasing areas taken with STM at 77 K (first and second columns, color scale in units of \%), strain histogram (third column) of the strain map shown in the first column, and strain histogram (fourth column) obtained from Raman spectroscopic studies of different areas of multiple PECVD-graphene samples grown on Cu foils. {\bf (b)} From left to right, compression/dilation strain maps over successively decreasing areas taken with STM at 77 K (first and second columns, color scale in units of \%), strain histogram (third column) of the strain map shown in the first column, and strain histogram (fourth column) obtained from Raman spectroscopic studies of different areas of multiple thermal CVD-grown graphene samples on Cu foils. The strain obtained from STM topography is largely consistent with the findings from Raman spectroscopic studies, both showing significantly reduced strain in the PECVD-grown graphene.}
\label{fig4}
\end{figure}

Having demonstrated the effect of strain on the electronic properties of graphene and our ability of developing nearly strain-free graphene, our next step is to employ theoretical simulations to assist our empirical designs of nanostructures on substrates in order to induce the desirable strain for graphene overlay on these substrates. 

\section{Theoretical simulations of strain-induced psuedo-magnetic fields}

We employed molecular dynamics (MD) techniques to compute the spatial distributions of the displacement field, strain tensor and pseudo-magnetic field for given engineered nanoscale structural distortions to graphene. For simplicity, we chose either a nano-sphere or a nano-hemisphere with a varying radius as the building block for constructing different nanostructures on the substrate. 

\subsection{Methods of the simulations}

For the MD simulations, we used the software package LAMMPS, which is available on the website http://lammps.sandia.gov. We considered a monolayer squared-shape graphene sheet with a fixed number of monolayer carbon atoms, and assumed that the positions of the carbon atoms at the boundaries of this graphene sheet remained invariant throughout the simulations. To induce structural distortions, we moved the graphene sheet adiabatically towards either a nano-sphere or a nano-hemisphere of gold nanoparticle until a desirable maximum height $h_0$ relative to the boundaries of the graphene sheet was reached (Fig.~5a-b), and then relaxed the entire system until it reached equilibrium. Here the load necessary to move the graphene sheet towards the nanostructure was comparable to the combined effect of ambient air pressure and gravity. 

Next, to ensure that the intrinsic properties of graphene were largely preserved without significant perturbations from the nanostructured substrate, we assumed that the attractive coupling between the graphene sheet and the underlying nanostructured substrate was sufficiently weak that it did not directly affect the graphene Hamiltonian. On the other hand, the attractive interaction must also be sufficiently strong to ensure proper conformation of graphene to the nanostructured substrate. We found that this situation could be realized by inserting a monolayer of hexagonal boron nitride (h-BN) in-between the graphene sheet and the nanostructure/substrate, provided that the crystalline structure of the h-BN layer was aligned at an incommensurate angle relative to the graphene sheet so that no superlattice structures formed within the finite sheet of graphene under consideration. 

The rationale for the choice of h-BN for the substrate material was based on the empirical fact that graphene on h-BN exhibited as excellent electrical mobility and fractional quantum Hall effects as those of suspended pristine graphene, in sharp contrast to substantially degraded mobility of graphene on other substrates (such as SiO$_2$/Si). Therefore, we employed realistic parameters for the van der Waals interaction of BN with carbon atoms and assumed isothermal conditions throughout the simulations described in Section 3. We also applied h-BN to all nanostructured substrates for empirical strain-engineering of graphene to be discussed in Section 4.         

Based on the approach outlined above, we were able to locate the three-dimensional coordinates of all carbon atoms on the graphene sheet and determined the two-dimensional coordinates of the three-dimensional displacement fields ($u_x$, $u_y$, $h$) and the resulting strain tensor components ($u_{xx}$, $u_{yy}$, $u_{xy}$) from Eq.~(\ref{eq:Strain}). Finally, we computed the vector potential components ($A_x$, $A_y$) from Eq.~(\ref{eq:Gauge}) and the pseudo-magnetic field components ($B_x$, $B_y$). 

\subsection{Strain induced by an isolated nanosphere on the substrate for graphene}

Using the aforementioned criteria and Eq.~(4), we obtained in Fig.~5c-f maps of the strain tensor components $u_{xx}$, $u_{yy}$, $u_{xy}$ and the resulting pseudo-magnetic field $B_s$ for a graphene sheet of $(100 \times 100)$ unit cells, which was $(24.6 \times 24.6)$ nm$^2$ in area, under the distortion of a nanoparticle with a diameter $d = 2.4$ nm and a maximum height $h_0 = 2.4$ nm.  

\begin{figure}
\centering
\includegraphics[width=3.4in]{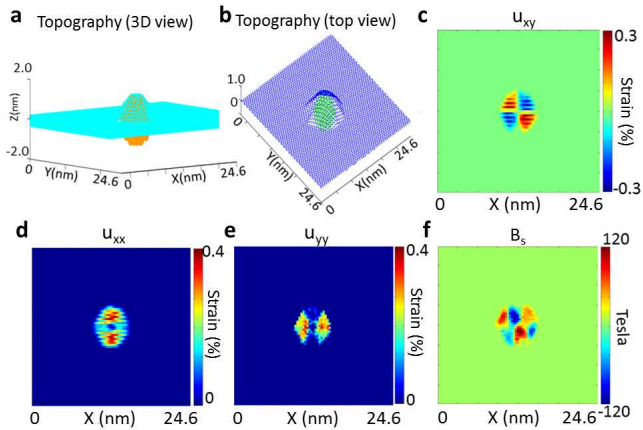}
\caption{MD simulations of the strain effects on a $(24.6 \times 24.6)$ nm$^2$ graphene sheet induced by a nanoparticle with a diameter 2.4 nm and a maximum height $h_0 = 2.4$ nm: {\bf (a)} A three-dimensional illustration of the structural distortion to the graphene sheet above a nanoparticle. {\bf (b)} The two-dimensional topographic distortion of the graphene sheet shown in (a). {\bf (c)} Spatial map of the strain tensor $u_{xy}$. {\bf (d)} Spatial map of the strain tensor $u_{xx}$. {\bf (e)} Spatial map of the strain tensor $u_{yy}$. {\bf (f)} Spatial map of the pseudo-magnetic field $B_s$.}
\label{fig5}
\end{figure}

It is interesting to note that the spatial distribution of the strain-induced pseudo-magnetic fields $B_s$ reveals an approximate three-fold symmetry pattern with alternating polarity of $B_s$ values that maximize around the location of the gold nanoparticle. The orientation of the three-fold symmetry pattern coincides with the zigzag direction of the graphene lattice structure, suggesting that the alignment of the nanoscale structural distortions relative to the crystalline lattice of graphene is an important parameter in the strain engineering of graphene. We further note the large magnitude of $B_s$ values (up to $\sim 120$ Tesla, Fig.~5f), suggesting substantial modifications of the graphene electronic properties under such a nanostructure of comparable lateral and vertical dimensions. For comparison, if the maximum height $h_0$ is reduced from 2.4 nm to 1.2 nm, the maximum $|B_s|$ value decreases from $\sim 120$ to $\sim 77$ Tesla, which is consistent with the strong dependence of the induced strain on the aspect ratio of the building block.    

\subsection{Theoretical results due to strain induced by two identical nanostructures on the substrate}

In order to better understand the effects of nanostructure alignment relative to the graphene crystalline lattice and the correlation of strain induced by  separate nanostructures, we examined the spatial maps of $B_s$ due to two identical nanospheres and two identical nano-hemispheres aligned relative to either the zigzag or the armchair direction and under varying separations $D_0$ between the centers of the two nanoparticles. As shown in Fig.~6a-b, for two hemispheres of $d = 2.4$ nm diameter and $h_0 = 1.2$ nm, the resulting pseudo-magnetic field for particle alignment along the zigzag direction appears to exhibit stronger correlation than along the armchair direction. In particular, for the nanoparticle separation $D_0 \le 2d$, nearly uniform stripes of pseudo-magnetic fields appeared along the zigzag direction, with alternating polarities of maximal $|B_s|$ values extending over nearly equal widths ($\sim 1$ nm) perpendicular to the zigzag direction (Fig.~6a). In contrast, insignificant spatial correlation was found for the nano-hemispheres aligned along the armchair direction for all appreciable $D_0$ values, as shown in Fig.~6b. 

For comparison, the $B_s$ distributions induced by two nanospheres (Fig.~6c-d) with $h_0 = 2.4$ nm also revealed stronger spatial correlation along the zigzag direction. However, in contrast to the nearly equal widths of alternating pseudo-field polarities in Fig.~6a with $h_0 = 1.2$ nm, the widths of alternating polarities of maximal $|B_s|$ values are not equally spaced so that substantial regions of the strained areas are covered by relatively smaller $|B_s|$ values. On the other hand, the spatial correlation of the pseudo-magnetic fields between the two nanoparticles appears to be more extended for $h_0 = 2.4$ nm than that for $h_0 = 1.2$ nm, suggesting that a larger degree of z-axis distortion in graphene also induces a longer range of lateral strain.

\begin{figure}
\centering
\includegraphics[width=3.4in]{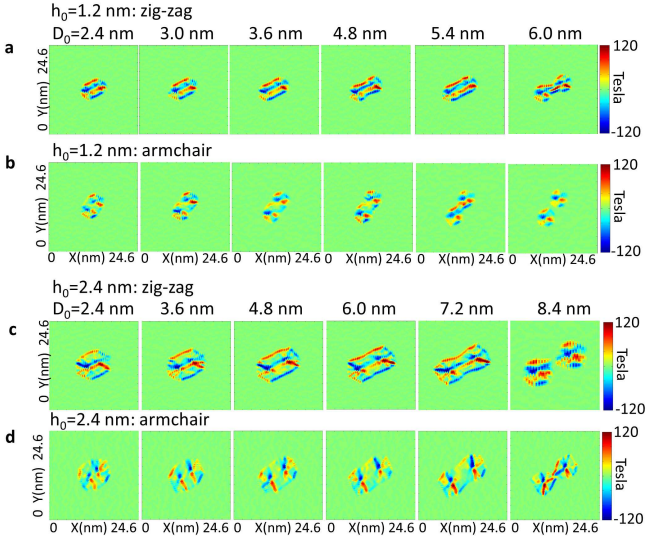}
\caption{MD simulations of the spatial maps of pseudo-magnetic fields $B_s$ over a $(24.6 \times 24.6)$ nm$^2$ graphene sheet induced by two nanoparticles with the same diameter 2.4 nm. {\bf (a)} Left to right: Two nano-particle aligned along the zigzag direction, with a maximum height $h_0 = 1.2$ nm and a separation $D_0$ = 2.4, 3.0, 3.6, 4.8, 5.4 and 6.0 nm, respectively. {\bf (b)} Left to right: Two nano-particle aligned along the armchair direction, with a maximum height $h_0 = 1.2$ nm and a separation $D_0$ = 2.4, 3.0, 3.6, 4.8, 5.4 and 6.0 nm, respectively. {\bf (c)} Left to right: Two nano-particle aligned along the zigzag direction, with a maximum height $h_0 = 2.4$ nm and a separation $D_0$ = 2.4, 3.6, 4.8, 6.0, 7.2 and 8.4 nm, respectively. {\bf (d)} Left to right: Two nano-particle aligned along the armchair direction, with a maximum height $h_0 = 2.4$ nm and a separation $D_0$ = 2.4, 3.6, 4.8, 6.0, 7.2 and 8.4 nm, respectively.}
\label{fig6}
\end{figure}

\subsection{General strategies for simulating arrays of nano-structures}

In addition to linear alignments of the building blocks of nano-spheres and nano-hemispheres, we examined different arrangements of the building blocks, such as three nano-spheres arranged in an equal triangular shape shown in Fig.~7a and 7b for each two nano-particles aligned along the zigzag and armchair directions, respectively; and for four nano-particles arranged in a squared shape shown in Fig.~7c for each two nano-particle aligned along alternating zigzag and armchair directions. In contrast to the findings in Fig.~6 where significant differences are apparent between nanostructures along the zigzag and the armchair directions, both types of triangular arrangements in Fig.~7a and 7b appear to be quite similar, with one side of the triangular arrangement differing from the other two sides. This finding is analogous to the three-fold symmetry breaking situation of the Kekule distortion.~\cite{16} Furthermore, for the squared-shape arrangement of nanoparticles along alternating zigzag and armchair directions, there is no apparent correlation of the pseudo-magnetic field pattern along any one side of the square, suggesting a frustrated configuration.         

\begin{figure}
\centering
\includegraphics[width=3.4in]{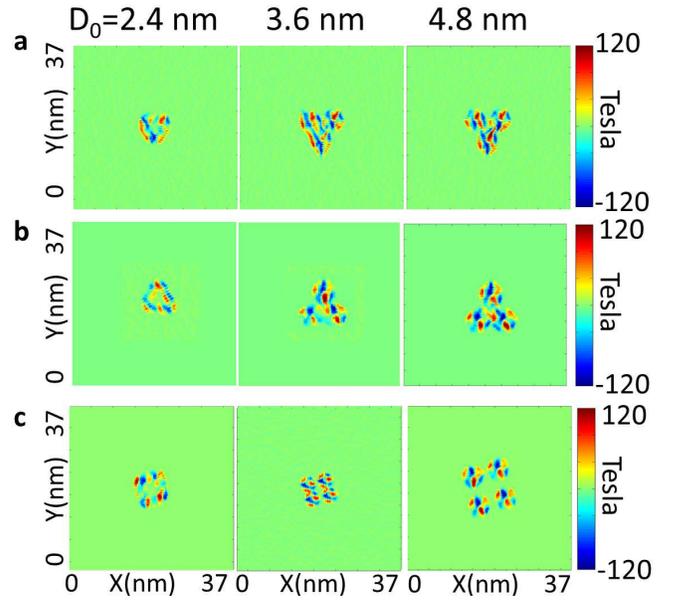}
\caption{MD simulations of the spatial maps of pseudo-magnetic fields $B_s$ over a $(37.0 \times 37.0)$ nm$^2$ graphene sheet induced by multiple nanoparticles with the same diameter 2.4 nm: {\bf (a)} Left to right: Three nano-particle aligned along the zigzag direction, with a maximum height $h_0 = 1.2$ nm and a separation $D_0$ = 2.4, 3.6 and 4.8 nm, respectively. {\bf (b)} Left to right: Three nano-particle aligned along the armchair direction, with a maximum height $h_0 = 1.2$ nm and a separation $D_0$ = 2.4, 3.6 and 4.8 nm, respectively. {\bf (c)} Left to right: Four nano-particle aligned along alternating zigzag and armchair directions, with a maximum height $h_0 = 1.2$ nm and a separation $D_0$ = 2.4, 3.6 and 4.8 nm, respectively.}
\label{fig7}
\end{figure}

Finally, we consider the pseudo-magnetic field distributions induced by ridge-like nanostructures aligned along either the zigzag or the armchair directions, as exemplified in Fig.~8a for the topography four nano-hemispheres of an identical diameter 2.4 nm. The maps of $B_s$ distributions over an area of $(37 \times 37)$ nm$^2$
for the nano-hemispheres aligned along the zigzag direction for inter-particle separations of $D_0 =$ 2.4, 3.6 and 4.8 nm are shown in Fig.~8b, and those for nano-hemispheres aligned along the armchair direction are illustrated in Fig.~8d. We find that for sufficiently closely spaced nano-hemispheres aligned along the zigzag direction, nearly uniform and stripe-like distributions of pseudo-magnetic fields could develop along the zigzag direction while the polarity of the pseudo-magnetic fields would alternate over $\sim 1$ nm width perpendicular to the zigzag direction. As exemplified in Fig.~8c for the height ($Z$) and the pseudo-magnetic field ($B_s$) distributions along the white line shown in the middle panel of Fig.~8b for $D_0 = 3.6$ nm, we find that the $B_s$ value correlates closely with $Z$ and maintains relatively large values ($\sim 100$ Tesla) except near the two ends of the nano-hemisphere distributions. In contrast, the pseudo-magnetic fields along the armchair direction tend to alternate in polarity more rapidly and so could not form extended regions of uniform pseudo-magnetic fields (Fig.~8d).  

\begin{figure}
\centering
\includegraphics[width=3.4in]{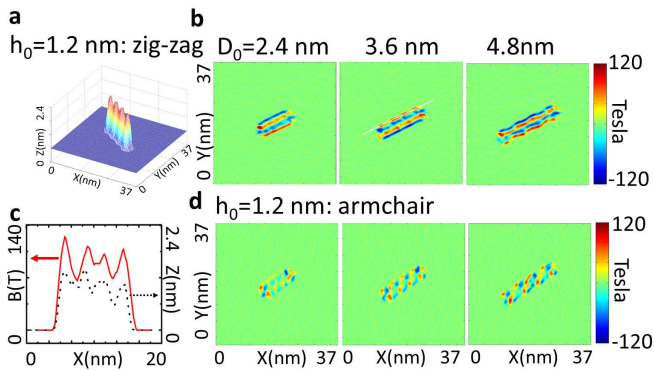}
\caption{MD simulations of the pseudo-magnetic field ($B_s$) distributions over a $(37 \times 37)$ nm$^2$ graphene sheet induced by four linearly aligned nano-hemisphere with the same diameter 2.4 nm and a maximum height $h_0 = 1.2$ nm : {\bf (a)} Topography of four nano-hemispheres aligned along the zigzag direction. {\bf (b)} Left to right: $B_s$ distributions of four nano-hemispheres aligned along the zigzag direction with a separation between consecutive nano-hemispheres being $D_0$ = 2.4, 3.6 and 4.8 nm, respectively. {\bf (c)} Comparison of the $B_s$ values along the zigzag direction (red solid curve) and the corresponding height variations (dotted black curve) for $D_0$ = 3.6 nm. {\bf (d)} Left to right: $B_s$ distributions of four nano-hemispheres aligned along the armchair direction and the separation $D_0$ = 2.4, 3.6 and 4.8 nm, respectively.}
\label{fig8}
\end{figure}

Based on the findings in Fig.~8, we expect that a ridge with a constant height and a constant slope would be able to produce relatively uniform $B_s$ values within an extended spatial width ($\sim 1$ nm) and alternating in sign across the lateral direction normal to the zigzag axis. 

\section{Experimental approaches to nano-scale strain engineering}

As discussed previously, our new PECVD approach to graphene growth can reproducibly fabricate nearly strain-free and large-area monolayer graphene with excellent properties, hence suitable for use in strain engineering. In this section we describe our empirical attempts to achieve controlled strain in initially strain-free graphene and demonstrate preliminary proof-of-concept results.  

\subsection{Engineering arrays of nanodots on silicon substrates by focused-ion-beam and electron-beam lithography}

Our first step towards nanoscale strain engineering of graphene was to investigate whether initially strain-free graphene could conform well to periodic arrays of nanostructures and would result in significant strain and pseudo-magnetic fields as theoretically predicted. To this end, we fabricated periodic spherical nanostructures using a dual focused ion beam (FIB) and scanning electron microscope (SEM) (FEI Nova NanoLab™ 600 DualBeam). The primary Ga$^+$ ion beam was operated at 30 keV and beam current was 10 pA. After the fabrication, we determined the topography of these nanostructures by means of both SEM and atomic force microscopy (AFM). The SEM images were taken with 5 kV acceleration voltage, 98 pA beam current, and a working distance $\sim 5$ mm. The AFM images were acquired in the tapping mode by Bruker Dimension Icon AFM. 

The nanostructures that we fabricated on silicon using the Ga-FIB consisted of nanodots with a diameter of $\sim 220$ nm and a height of $\sim 60$ nm, as shown in Fig.~9a. These nanostructures were repeated into $(4 \times 4)$ arrays with an average separation of $\sim 440$ nm (between the centers of two neighboring nanodots) within a $(2.5 \times 2.5) \ \rm \mu m^2$ square, and the same arrangements were repeated over a $(20 \times 20) \ \rm \mu m^2$ area. Monolayer hexagonal boron nitride (h-BN) was then transferred onto the Si-nanostructures, followed by the transfer of PECVD-graphene onto the substrate of BN/Si-nanostructures. We performed AFM and SEM measurements on every step of the process, and found that BN conformed very well to the Si-nanostructures, as exemplified by the AFM image in Fig.~9b. Interestingly, upon the deposition of graphene on BN/Si-nanostructures, we found that graphene appeared to wrinkle up slightly along the nanostructures, as exemplified by the AFM image in Fig.~9c. This situation is analogous to our findings from the MD simulations, which revealed the appearance of graphene wrinkles along the alignment of its underlying nanostructures if the nanostructures were not too far apart. 

Another approach to construct regular arrays of nanostructures was the use of electron-beam lithography. Specifically, we wrote regular arrays of nanodot sturctures on silicon substrates coated with PMMA. After development and metal deposition, the PMMA was lift-off, thus forming arrays of gold nanodots, as exemplified in Fig.~9d-e for a typical nanodot diameter and inter-dot separation of $\sim 52.5$ nm. The advantage of using electron-beam lithography to Ga-FIB is the ability of making a much larger area of the same density of the nanostructures by using masks, which is in contrast to the slow process of writing individual nanodots by the FIB. On the other hand, special care must be taken to ensure that gold nanoparticles remain in position throughout the placement of h-BN and graphene.     

\begin{figure}
\centering
\includegraphics[width=3.4in]{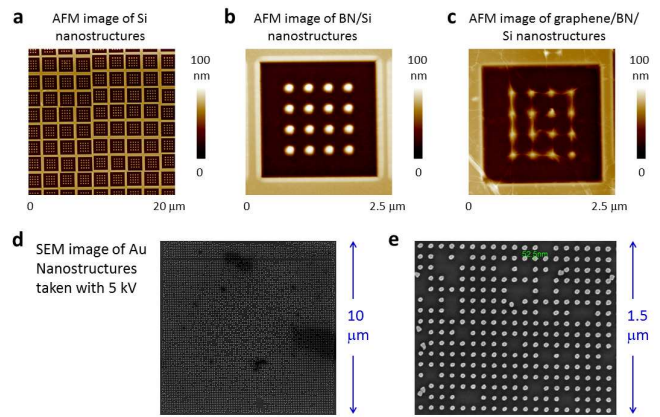}
\caption{AFM images of nanostructures for strain engineering of graphene: {\bf (a)} Arrays of Si-nanostructures created by Ga-FIB over a $(25 \times 25) \ \rm \mu m^2$ area. The diameter and height of the nanostructures were $\sim 220$ nm and $\sim 60$ nm, respectively, and the inter-dot separation between the centers of the neighboring nanodots was $\sim 440$ nm. {\bf (b)} One monolayer of BN on top of Si-nanostructures over a $(2.5 \times 2.5) \ \rm \mu m^2$ area, showing excellent conformation of BN to the Si nanodots. {\bf (c)} One monolayer of graphene on top of the structure shown in (b), revealing graphene ``wrinkles'' along the aligned Si nanodots. {\bf (d)} Arrays of gold nanoparticles on a silicon substrate over a $(10 \times 10) \rm \mu m^2$ area, created by electron-beam lithography. {\bf (e)} A magnified image of the $(2 \times 2) \rm \mu m^2$ boxed area shown in (d), revealing mostly regular gold nanoparticles with a diameter and an inter-particle separation of $\sim 52.5$ nm on the silicon substrate except for several displaced nanoparticles.}
\label{fig9}
\end{figure}

\subsection{Raman spectroscopic studies of the nanostructure-induced strain effects on graphene}

To investigate the macroscopic strain induced by regular arrays of nanostructures, we performed spatially resolved Raman spectroscopic studies on the graphene/BN/Si-nanostructure sample shown in Fig.~9c, and also compared the spectra with those taken from reference areas of the same sheet of graphene without underlying Si-nanostructures. The Raman spectrometer used a laser excitation of 532 nm with a spot size of about 500 nm. We mapped at 0.5 $\rm \mu m$ per pixel steps. Each 2D-band spectrum was fit to a single Lorentzian, and the frequency peak position was assigned as the corresponding 2D-band frequency. While the spatial resolution of the Raman spectrometer at 0.5 $\rm \mu m$ was not sufficient to resolve the spatial distribution of strain associated with the individual nanodots, it appeared that the strain effects induced by the larger squared features could be resolved in some places. Overall the spatial map of the 2D-band over the area of graphene above the nanostructures clearly revealed significant inhomogeneity (Fig.~10a), which was in sharp contrast to the spatial homogeneity of the 2D-band map of a controlled area of graphene on top of a flat region of the BN/Si substrate (Fig.~10b). A more quantitative comparison of the spatial distributions of the 2D-band frequency in these two areas is illustrated by the histograms in Fig.~10c, where a much broader 2D-band and therefore much wider strain variations induced by the presence of nanostructures is clearly demonstrated.  

\begin{figure}
\centering
\includegraphics[width=3.4in]{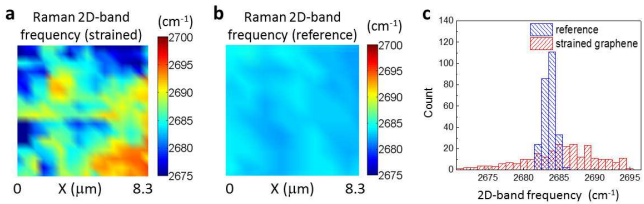}
\caption{Raman spectroscopic studies of the spatial distributions of strain in graphene: {\bf (a)} Spatial map of the 2D-band of graphene on top of a $(10 \times 10) \rm \mu m^2$ area of nanostructure arrays shown in Fig.~9a. The pixel size of the Raman map is $0.5 \rm \mu m$, which is insufficient to resolve the small nanostructures. Nonetheless, the map still reveals strong spatial inhomogeneity, with some of the highly strained lines consistent with the larger patterned features of the squares. {\bf (b)} Spatial map of the 2D-band of graphene over a $(10 \times 10) \rm \mu m^2$ reference area above a flat region of the BN/Si substrate. The pixel size of the Raman map is $0.5 \rm \mu m$, and the map appears to be spatially homogeneous. {\bf (c)} Comparison of the histogram of the 2D-band between strained graphene in (a) and the strain-free reference area in (b), showing much broader 2D-band distributions in the strained area.}
\label{fig10}
\end{figure}

\subsection{Engineering nanodots on silicon by self-assembly of gold nanoparticles}

In order to investigate the strain distributions with much higher spatial resolution, STM must be employed. However, it is generally very challenging to locate the small $(20 \times 20) \ \rm \mu m^2$ patterned area within a relatively large sample area, typically on the order of $(5 \times 5) \ \rm mm^2$, under the atomically sharp STM tip. While this issue may be addressed by enlarging the total area of patterned nanostructures, another plausible approach was to distribute metallic nanoparticles over the entire substrate by means of self-assembly. This approach could enable quasi-periodic distributions of nanoparticles with a limited range of diameters. Although not ideal for well controlled strain engineering, the use of self-assembled nanoparticles in place of either FIB or electron-beam fabricated nanostructures could provide preliminary and semi-quantitative verifications for our theoretical designs, and so was worthy of exploration.

For this work, the self-assembled gold nanoparticles were developed from solutions on silicon substrates using block copolymer lithography (BCPL)~\cite{24}. The preparation procedure is briefly summarized below. 

The solutions were mixed in a Pyrex 5 ml micro volumetric flask that was specifically designed for microchemical work. The glassware was cleaned in aqua regia, rinsed in DI water, followed by ultrasonication, and triple rinsed in DI water followed by mild baking to remove any residual moisture. The copolymer (Polymer Source, Inc.) was the diblock copolymers 25.5 mg of polystyrene (81,000)-block-poly(2-vinylpyridine)(14,200), which was dissolved in 5 ml of ultra-high purity toluene (Omnisolve, 99.9\%) and spun vigorously with a Teflon stir bar for four days. The solution was clear. The gold precursor was Gold(III) chloride hydrate (99.999\% trace metals basis Aldrich, Inc.), and 14.6 mg were added to the diblock copolymer solution under dry nitrogen and in a darkened environment. The solution was then stirred vigorously for l0 days to insure particle uniformity. The solution remained clear upon stirring, but the color changed to amber.

The gold arrays were formed on mechanical grade silicon (5 x 5) mm without removing the native oxide. The silicon substrate was first blown with dry nitrogen to remove particulates and then subjected to a 100 W oxygen plasma (Technics Inc., Planar Etch) for five minutes to remove any hydrocarbon residue. The substrate was loaded on a spin-coater and held in place with a vacuum chuck. The Au-BCPL solution was dropped onto the silicon substrate so that the entire top surface was covered. The sample remained covered for 30 s before starting the spin process. Samples were spun at 3000 RPM for 60 s. The substrate was removed from the spin-coater and placed in an oxygen plasma for 25 minutes at 100 W, and was subsequently removed from the plasma and baked overnight at 220 $^{\circ}$C.

Upon drying the solution and removing the polymer, quasi-periodic gold nanoparticles of diameters ranging from 14 to 20 nm remained and covered the surface of the silicon substrate, with an average separation between adjacent nanoparticles ranging from 10 to 30 nm, as exemplified in Fig.~11a for an AFM image over an area of $(1 \times 1) \ \rm \mu m^2$. Next, a monolayer BN was placed over the Au-nanoparticles/silicon substrate, which was found to conform well to the nanoparticles, as shown in Fig.~11b. Finally, a monolayer PECVD-grown graphene was placed on top of the BN/Au-nanoparticles/silicon, which induced wrinkles on graphene, as illustrated in Fig.~11c and similar to the findings in Fig.~9c. In particular, we note a general preferential wrinkle alignment along approximately 150$^{\circ}$ direction relative to the x-axis of the plot. This preferential direction might be along the zigzag direction of the graphene lattice based on our MD simulations, although further atomically resolved STM studies of the wrinkle structures will be necessary to verify this conjecture.

\begin{figure}
\centering
\includegraphics[width=3.4in]{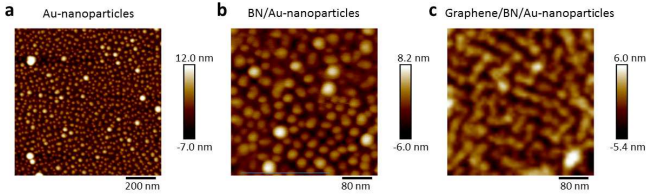}
\caption{AFM images of self-assembled nanostructures: {\bf (a)} Quasi-periodic gold nanoparticles of diameters $14 \sim 20$ nm on silicon over an area of $(1 \times 1) \ \rm \mu m^2$. {\bf (b)} One monolayer of h-BN on top of the same quasi-periodic gold nanoparticles in (a) over a $(500 \times 500) \ \rm nm^2$ area, showing excellent conformation of BN to the Au nanoparticles. {\bf (c)} One monolayer of PECVD-grown graphene on top of the structure shown in (b), revealing graphene ``wrinkles'' with a preferential wrinkle alignment along approximately 150$^{\circ}$ direction relative to the x-axis of the plot.}
\label{fig11}
\end{figure}

\subsection{Characterizations of strain-induced modifications to the density of states by scanning tunneling spectroscopy}

We performed STM and STS studies on the aforementioned graphene/BN/Au-nanoparticle sample at 300 K to characterize the spatial distribution of strain. In Fig.~12a the STM topography over a $(140 \times 140) \ \rm nm^2$ area revealed the protrusion of graphene above nanoparticle structures. Close-up point spectroscopic studies around the nanoparticle structures (Fig.~12b, which corresponded to the lower left area of Fig.~12a) indicated that in the highly strained regions where maximum changes in the graphene height appeared (with $>\sim 10$ nm descent over a lateral dimension $<\sim 4$ nm), extra spectral features associated with the quantized density of states under a large pseudo-magnetic field appeared on top of the typical Dirac spectrum, as exemplified in Fig.~12c, where the excess quantized spectral features (Fig.~12d, after the subtraction of the Dirac spectrum) correspond to a pseudo-magnetic field $B_s \sim 55$ Tesla according to Eq.~(\ref{eq:En}). In contrast, for relaxed regions of the graphene sample, the tunneling spectra recovered the standard Dirac spectrum under finite thermal smearing (Fig.~12e). Interestingly, the magnitude of the strain-induced pseudo-magnetic field is somewhat smaller than the result of $B_s \sim 77$ from the MD simulations for an isolated nano-hemisphere with a diameter 2.4 nm and maximum height $h_0 = 1.2$ nm, the latter having an aspect ratio comparable to that of the Au-nanoparticles used in this work. This finding is reasonable because generally for the same aspect ratio of structural distortions, the induced strain decreases gradually with the increasing physical size of the distortion.   

\begin{figure}
\centering
\includegraphics[width=3.4in]{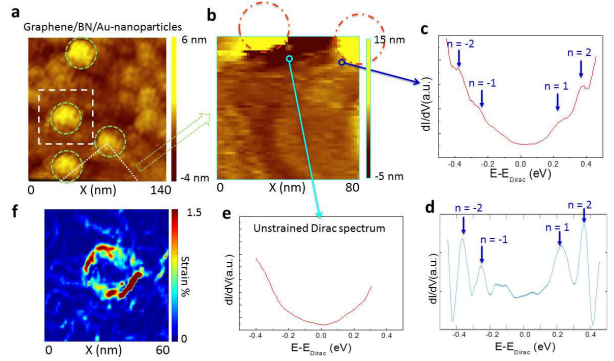}
\caption{STM studies of the topography and tunneling spectroscopy in graphene/BN/Au-nanoparticles at 300 K: {\bf (a)} Surface topography of a $(140 \times 140) \rm nm^2$ area, showing surface protrusion above Au-nanoparticles of $\sim 25$ nm diameter. {\bf (b)} Surface topography of a $(80 \times 80) \rm nm^2$ area that corresponded to the region partially indicated by the white box in (a), showing two partial nanoparticles in the upper region. {\bf (c)} Point spectrum taken at the location indicated by the blue circle on the right, which corresponded to a region of rapid changes in height. Excess enhancement in the tunneling conductance appeared at quantized energies that corresponded to Landau levels associated with a pseudo-magnetic field on the order of $B_s \sim 55$ Tesla according to Eq.~(\ref{eq:En}). {\bf (d)} The same point spectrum as in (c) after subtraction of the background Dirac spectrum. {\bf (e)} Point spectrum taken at the location indicated by the light blue circle on the left, which corresponded to a flat region of negligible strain so that the spectrum is consistent with the standard Dirac spectrum. {\bf (f)} The magnitude of the biaxial strain map of graphene obtained from atomically resolved topographic studies of the $(60 \times 60) \rm nm^2$ area indicated by the white dashed box in (a), showing maximal strain near the periphery and minimal strain around the top of the nanoparticle.}
\label{fig12}
\end{figure}

\section{Discussion}

Although the concept of nanoscale strain engineering of initially strain-free graphene samples has been verified semi-quantitatively based on the aforementioned studies, a number of challenges remain. 

First, as demonstrated by the MD simulations, the pattern and strength of the pseudo-magnetic field depend strongly on the sizes, shapes, orientations and patterns of the nanostructures on the substrate for graphene. While modern technology for nano-fabrication could manage most of the requirements for the nanostructures, the necessity to align the graphene lattice with high accuracy in orientation relative to the nanostructures is a nontrivial task.   

Second, any finite interactions between the substrate material and graphene would complicate the effect induced by generic structural distortions. While our insertion of a monolayer of BN between the substrate and graphene was intended to minimize the influence of the substrate and to preserve the generic properties of graphene, in the event of nearly perfect alignment of graphene with the underlying h-BN lattices, the Dirac electrons of graphene could become gapped~\cite{20} so that the theoretical foundation for strain engineering of gapless Dirac fermions would no longer hold. 

Third, our MD simulations have assumed perfectly local pseudo-magnetic fields in response to the local strain. This assumption is justifiable if the carrier density in the graphene sheet is sufficiently low so that electronic screening effects are negligible. On the other hand, doping effects of spatially inhomogeneous charged impurities could result in weakened pseudo-magnetic fields and broadened Landau levels, which may account for the relatively broad linewidths observed in the tunneling spectra in Fig.~12c.    

Quantitatively, we may incorporate the non-local correction to the evaluation of an effective pseudo-magnetic field $\langle B_s \rangle$ at position $(x_0, y_0)$ by following formula: 
\begin{eqnarray}
\langle B_s (x_0, y_0) \rangle &= (2 \pi \ell _0 ^2 )^{-1} \int _{-X} ^X dx \int _{-Y} ^Y dy \ B_s (x,y) \qquad \qquad \qquad \nonumber\\
& \times \exp \lbrack - \sqrt{(x-x_0)^2+(y-y_0)^2}/ \ell _0 \rbrack, 
\label{eq:meanBs}
\end{eqnarray}
where $\ell _0$ denotes the mean free path, and we have assumed that the x-range (y-range) of the sample expands from $-X$ to $X$ (from $-Y$ to $Y$). For a constant pseudo-field distribution over an infinite sample, we have $\langle B_s (x,y) \rangle = B_s (x,y)$ everywhere within the sample as expected. On the other hand, a short mean free path $\ell _0$ would result in effective $\langle B_s (x,y) \rangle$ values significantly smaller than the local value of $B_s (x,y)$ according to Eq.~(\ref{eq:meanBs}).   

We further remark that the relevant characteristic length involved in the non-local effect of pseudo-magnetic fields in Eq.~(8) should be the mean free path $\ell _0$ rather than the pseudo-magnetic field length $\ell _B$, because the contributions from the strain-induced gauge potential are only perturbative to the total Hamiltonian of the Dirac fermions, hence not all Dirac fermions are completely localized to the length scale of $\ell _B$ by the presence of a pseudo-magnetic field, which is different from the situation for a global time-reversal symmetry breaking magnetic field. 

For a given spatial distribution of $B_s(x,y)$, if we define the deviation of the effective pseudo-magnetic field $\langle B_s (x,y) \rangle$ from its local value by $\delta B_s$, we find that from Eq.~(\ref{eq:En}), the linewidth $\delta E_n$ of the Landau level energy $E_n$ becomes $\delta E_n \propto \sqrt{|n/B_s|} \delta B_s$. Therefore, the linewidth of $E_n$ increases with increasing non-local corrections to $B_s$. On the other hand, for a given pseudo-magnetic field distribution we expect sharper Landau levels at locations with larger $B_s$ values. Ultimately, we expect much sharper conductance peaks for pseudo-magnetic fields induced on nearly impurity-free graphene by well ordered and well shaped nanostructures, and hope to verify this notion by further experimental investigation.     

Finally, we note that the presence of pseudo-magnetic fields can lift the degeneracy of the two valleys, which could enable valleytronic applications. As exemplified in Fig.~13, we can in principle detect the valley Hall effect (VHE) in a design similar to that depicted for measurements of the VHE on the graphene/h-BN superlattice structure,~\cite{20} except that we do not align the crystalline lattices of graphene and h-BN. Rather, we would use a tetrahedron-shaped nanodot to split the valley currents and employ a backgate to tune the Fermi level to the Dirac point for maximum response. This design is equivalent to a valley field-effect-transistor enabled by the strain-induced topological currents. Moreover, the ability to split the valley degeneracy into two paths with opposite chirality may be utilized in novel designs of either passive devices such as chiral electronics for detectors and light polarizers, or active devices such as tabletop free electron lasers with strain-induced, polarity-alternating pseudo-magnetic fields (see, for example, Fig.~8b) as the undulators for accelerating valley-polarized Dirac fermions into synchrotron radiation.    

\begin{figure}
\centering
\includegraphics[width=3.4in]{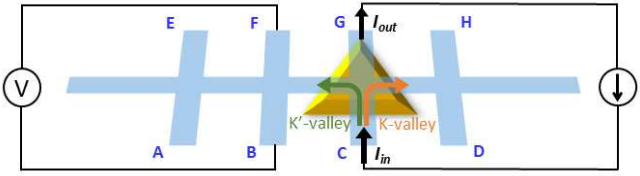}
\caption{Electrical circuit design for detecting the valley Hall effect (VHE) in graphene under strain-induced pseudo-magnetic fields: The blue-shaded area corresponds to patterned graphene in a Hall bar geometry. The green feature represents a tetrahedron-shaped nanostructure for straining the graphene overlay on it. The strained region would induce a spatially varying pseudo-magnetic field with opposite signs to the Dirac electrons in the $K$ and $K^{\prime}$-valleys, so that a transverse momentum can be provided to deflect electrical currents of opposite pseudo-spins to different directions over the strained area. Therefore, in additional to the longitudinal resistance ($R_{xx}$) that can be determined from $R_{xx} = (V_{BC}/I_{AD})$, a non-local resistance ($R_{NL}$) can be measured from $R_{NL} = (V_{BF}/I_{CG})$ as shown above, or equivalently from $(V_{DH}/I_{CG})$. By placing the graphene Hall bar on the h-BN/SiO$_2$/Si substrate and attaching a back gate to the Si, the Fermi level of the graphene can be controlled relative to the Dirac point by tuning the gate voltage ($V_G$) so that a sharp peak in $R_{NL}$-vs.-$V_G$ is expected when the Fermi level coincides with the Dirac point.~\cite{20} This configuration is therefore a field effect transistor. We remark that the distance between CG and BF must be less than the mean free path $\ell_0$ to ensure optimized valley splitting if only one tetrahedron-shaped nanostructure is constructed. In the event of smaller $\ell_0$ values, multiple tetrahedron-shaped nanostructures that space $\sim \ell_0$ apart from each other may be aligned between CG and BF to prolong the spatial extent of the valley splitting effects.}
\label{fig13}
\end{figure}

\section{Conclusion and Outlook}

We have reviewed in this work the theoretical foundation for nanoscale strain engineering of graphene, and have demonstrated the use of molecular dynamics techniques for designing various nanostructures to realize different patterns of pseudo-magnetic fields. We have also presented experimental evidences for strain-induced charging effects and giant pseudo-magnetic fields, and described feasible empirical approaches based on nano-fabrication techniques to realizing strain-engineering of pseudo-magnetic fields and valleytronics.      

While the nanoscale strain engineering effects may be determined by STM, we emphasize that useful designs must ensure that the strain effects are extended to mesoscopic and even macroscopic scales for realistic device applications. Hence, the use of theoretical simulations (such as the molecular dynamics techniques) to assist the design of collective nanostructures for desirable strain distributions and device performance can greatly improve the effectiveness of experimental implementations. 

Finally, we remark that the feasibility of nanoscale strain-engineering and valleytronics in graphene is based on the unique electronic properties of this novel nano-material and the availability of modern nanotechnology. With due ingenuity, imagination and diligence, the interesting interplay of structural, electronic and topological properties of nano- and meta-materials is likely to open up new frontiers in scientific and technological exploration that have not been envisioned before.        
\\
\\

\begin{acknowledgments}
This project was jointly supported by the National Science Foundation under the Institute for Quantum Information and Matter (IQIM) at California Institute of Technology, a grant from the Northrup Grumman Cooperation, and a gift from Mr. Lewis van Amerongen.
\end{acknowledgments}


\end{document}